\pdfoutput=1
\documentclass[aps,pra,twocolumn,reprint,10pt]{revtex4-1}

\usepackage{bm,amsmath,amssymb,graphicx,url,hyperref}

\newcommand{\bs}{\boldsymbol}
\newcommand{\re}{\text{Re}}
\newcommand{\im}{\text{Im}}

\begin{document}

\title{The connection between polarization calculus and four-dimensional rotations}
\author{Magnus Karlsson}
\email{magnus.karlsson@chalmers.se}

\affiliation{Photonics Laboratory, Department of Microtechnology and Nanoscience\\ Chalmers University of Technology SE-41296 Gothenburg, Sweden}
\begin{abstract}
We review the well-known polarization optics matrix methods, i.e., Jones and Stokes-Mueller calculus, and show how they can be formulated in terms of four-dimensional (4d) rotations of the four independent electromagnetic field quadratures. Since 4d rotations is a richer description than the conventional Jones and Stokes-Mueller calculi, having six rather than four degrees of freedom (DOF), we propose an extension of those calculi to handle all six DOF. 

For the Stokes-Mueller analysis, this leads to a novel and potentially useful extension that accounts for the absolute phase of the optical field, and which can be valuable in the areas where the optical phase is of interest, e.g. interferometry or coherent communications. As examples of the usefulness we use the formalism to explain the Pancharatnam phase by parallel transport, and shows its connection with the Berry phase.

In addition we show that the two extra DOF in the 4d description represents unphysical transformations, forbidden for propagating photons, since they will not obey the fundamental quantum mechanical boson commutation relations. 
\end{abstract}

\pacs{42.25.Ja	Polarization}

\maketitle

\section{Introduction}
This paper presents an new formalism for polarization calculus by the use of four dimensional (4d) rotations. This will then lead to extension of the classical polarization-optics calculi (due to Jones and Stokes-Mueller, respectively), to be as geometrically rich as the set of four-dimensional rotations. We begin with an historical overview that will lead to a motivation and outline of this paper.

\subsection{Historical background}
The use of matrix methods to calculate the polarization evolution for electromagnetic waves was, given the earlier discoveries within optics, a relatively late invention, originally proposed by Jones \cite{jones1,*jones2,*jones3,*jones4,*jones5,*jones6,*jones7,*jones8} and Mueller \cite{mueller43,parke48} in the 1940:ies. 
One reason was probably that before this time relatively few polarization devices were available, and there was no demand for what we now refer to as ``Jones calculus'' and ``Stokes-Mueller calculus''. However all the knowledge was there: already in 1818 Young (following a proposition by Hooke in 1757, and also work by Arago in the early 19th century) showed that light was a transverse wave. Indeed both Maxwell \cite[Ch. 21, Par. 812-813] {maxwell91} and Poincar\'e \cite{poincare92} treated polarization as two independent transverse field components with orthogonal directions in their books from the early 1890s. Poincar\'e even showed that the complex ratio of these two waves could be conveniently mapped  (via a stereographic projection) to the sphere that since then bears his name \cite[Ch. 12]{poincare92}. In \cite{jerrard54} a more modern discussion of this projection is given. In fact, Poincar\'e even showed how to propagate the complex amplitude ratio of the two field components via a bilinear transformation \cite[Ch. 12, Par. 158]{poincare92}, which was an early precursor to today's Jones calculus.

Paralleling those developments, the 1852 landmark paper by Stokes \cite{stokes52} is worth mentioning for two reasons; (i) it created the first framework for analyzing polarized and unpolarized light (leading to the first mathematical description of the Fresnel-Arago interference laws \cite[p. xv]{collett93}), and (ii) for the introduction of the four parameters (nowadays referred to as \emph{Stokes parameters}, which can be formed in to a 4-component \emph{Stokes vector}) that enabled this framework. The Stokes parameters, which originated from experiments and were based on essentially measurable quantities, seemed in the late 19th century quite disconnected from the transverse field components of the electromagnetic field as it was described by Maxwell \cite{maxwell91} and Poincar\'e \cite{poincare92}. Later work in the first half of the 20th century, by e.g. Soleillet \cite{soleillet29}, Perrin \cite{perrin42}, and Mueller \cite{mueller43}, connected the Stokes parameters with the Cartesian coordinates of the Poincar\'e sphere, and also showed how the Stokes parameters are linearly transformed as light propagates through birefringent and polarizing media. The stage was thus set for the matrix methods of relating input to output states in polarization systems.

While Jones put forward the calculus named after him in a series of papers \cite{jones1}, the history of the Mueller calculus is a bit more involved. 
Mueller had originally described his matrix method in the (since 1960 declassified) defense report \cite{mueller43}, and then building on the earlier work by mainly Perrin \cite{perrin42}. A more thorough description was then given in Parke's 1948 PhD thesis \cite{parke48}, supervised by Mueller. 

The next important development was made by Falkoff and MacDonald \cite{Falkoff51}, who connected the Stokes parameters with the \emph{density matrix} (in classical optics also referred to as the \emph{coherency matrix}), which can be directly related to the Jones vector. In fact, this mathematical relationship was pointed out already 20 years earlier by Wiener  \cite[Ch. 9]{Wiener30}, although he was unaware of the physical relation to Stokes' four parameters. Important steps towards a more modern notation of polarization calculus were then made by Fano\cite{fano54} and Barakat\cite{barakat63}, who introduced the Pauli matrices as a convenient bridge between Jones vectors/coherency matrices and Stokes vectors. 

We thus have two ways of describing polarization transformations; via $2 \times 2$ complex Jones matrices, or the $4 \times 4$ real Mueller matrices. 
The Jones matrices has the advantage that they can describe also the absolute optical phase change. Often this is of little relevance, however interferometry and coherent optical communications are two notable and very important exceptions. The Mueller calculus has the advantage of also being able to describe partially polarized light. Since they describe the same phenomena, it should come as no surprise that there is a deeper, underlying mathematical relationship between Jones and Mueller matrices, first pointed out by Takenaka\cite{takenaka73a,*takenaka73b}. In the general case this stems from the known mathematical fact that the group of complex $2 \times 2$ Jones matrices is isomorphic to the Lorentz group of $4 \times 4$ transformations \cite{cloude86}. The fact that the Lorentz group is mostly associated with relativistic dynamics has sometimes obscured its role in Mueller calculus. Although the relationship have been discussed in the literature \cite{frigo86,cloude86,pellatfinet92,han97} it is not widely known in the optics community. 

Finally the use of \emph{quaternions} have been proposed for polarization calculus, both for the use as Jones (state) vectors \cite{pellatfinet89,*pellatfinet91} as well as to simplify the calculation of Jones matrix cascades \cite{karlsson04}. In this context, quaternion algebra can be used to replace, or computationally simplify, Jones or Mueller matrix multiplications. 

If we restrict the polarization transformations to pure birefringence (no polarization dependent losses), it is enough to use only 3 parameters of the Stokes vector. The Mueller matrices will then belong to the group of $3 \times 3$ rotation (orthogonal) matrices, and the subgroup with positive deteminants is commonly denoted  $O_3^+$. This group is isomorphic to the set of unitary $2 \times 2$ Jones matrices (the $SU(2)$ group) \cite{takenaka73b,cloude86}. An general introductory discussion on the $O_3^+$ - $SU(2)$ isomorphism can be found in \cite[Ch. 4.10]{arfken85}. Both these groups have 3 degrees of freedom (DOF), since 3 parameters are needed to describe a three-dimensional (3d) rotation (e.g. two for the axis and a third for the angle of rotation). In addition, for Jones matrices, the absolute phase can be added as a 4th degree of freedom.

In this work we will present a new description of electromagnetic wave transformations, namely that of using the four-dimensional real space. This might seem like a trivial extension of the Jones calculus - merely separating the real and imaginary parts of the complex (Jones-) field components into four real vector components.  However, as will be shown, and which is somewhat surprising, is that this leads to a \emph{more general description} than both the classical Jones and Stokes-Mueller formalisms.  

\subsection{Motivation and outline}
The description of a polarized electromagnetic wave in terms of its four real quadrature components was pioneered in the area of optical communications\cite{betti90,*betti91,cusani92}, where such a formalism is relevant since noise is (in the semi-classical model) linearly added independently to each quadrature, or dimension. The signal is then described by a real 4-component vector, with a length squared equal (or proportional) to the power. A lossless propagation of such a signal vector can then be described by the 4d real analog to the complex Jones matrix (see e.g. Eq. (9) in \cite{cusani92}), which has 4 independent parameters, or equivalently, DOF. Obviously, such a $4\times 4$ transfer matrix is a four-dimensional (4d) rotation, since it will not change the length (or power) of the state vector. However, 4d rotations form a 6-parameter group \cite{4drot}, implying that the 4 DOF Jones matrices does not cover the full set of possible transformations in real 4d space \cite{Karlsson10ecoc}.The question then emerges: these extra two DOF in the 4d rotations, what kind of transformations do they correspond to in Jones and Stokes space? This paper is addressing that question.

As we will see the answer is very interesting, both from a mathematical and physical perspective. There is also a real practical relevance from the application of matrix methods to fiber optic transmission. We will show that the two extra DOFs of the 4d rotations corresponds to \emph{forbidden transformations}, i.e. polarization transformations that can be mathematically described by extensions of Jones and Stokes-Mueller calculi, but which does not have any physical realization in electromagnetic wave propagation\footnote{at least known to me at the time of writing}. But then why are they of interest? They are of interest for at least three reasons: 
\paragraph*{Extension:} They enable us to extend the existing calculi to cover a richer set of dynamics. For example, we will present the (perhaps first) mathematically consistent generalization of the Stokes vector to account for the absolute phase of the optical field.
\paragraph*{Commutation:} The forbidden transformations commute with the allowed (conventional) ones. This is a unique property for these transformations that may have practical applications.
\paragraph*{Synthesis:} Even if the forbidden transformations do not exist in nature, they can be synthesized in digital signal processing and waveform generation and might as such be of use in e.g. coherent optical transmitters and receivers.

In this first paper on 4d polarization calculus we will limit the formalism to unitary (power conserving) transformations, which corresponds to purely birefringent media. It is quite likely possible, but beyond the present scope, to generalize this work to non-unitary transformations, as required for a description of e.g. polarization dependent losses. (N.B.: complex $2 \times 2$ Jones matrices have 8 DOF, real $4 \times 4$ matrices has twice as many, 16 DOF. What do the extra 8 DOFs correspond to?)

 It is also worth noting that even if the main application here is polarization calculus, there are numerous other areas within theoretical physics where the group theoretical properties of $SU(2)$ and $O_3^+$ are relevant, as. e.g. particle physics, 2-state quantum systems, spinor theory, or coupled mode theory. Thus the presented theory will certainly have wider applications than polarization calculus.
 
This article is organized as follows: In sections 2 and 3 we will for reference review the well-known unitary Jones and Stokes-Mueller calculi and how they can be mapped to each other. In section 4 we will describe the corresponding 4d rotations. In section 5 we describe the general (6 DOF) set of 4d rotations including the forbidden and allowed transformations, and their parameterization. Then in sections 6 and 7 we will connect back to the Jones and Stokes calculi, respectively, and see how they can be extended to deal with both sets of 4d rotations. We will show that the 6 DOF 4d rotation group form 2 commuting subgroups with 3 DOF each, of which the classical Mueller matrices correspond to one. We will also explicitly prove that the extra two DOF are forbidden, in the sense that they do not (contrary to the other four DOF) satisfy the fundamental boson commutation relations.  In section 8 we will, as an example, apply the generalized Stokes method to explain the Pancharatnam phase, and finally in section 9 conclude. The main results are summarized in Table \ref{newtable}. A number of appendices provide supporting material, definitions and details not necessary for the main understanding.

We employ the following notation conventions: Complex 2-vectors are denoted by boldface lowercase letters, as e.g. $\bs e$. Real 3-vectors (e.g. Stokes vectors or rotation vectors) are denoted with lowercase vector-arrow notation as $\vec e$. Unit vectors have a hat, as $\hat e$. Real 4-vectors are denoted with uppercase vector-arrow notation as $\vec E$. Matrices are usually uppercase, and their dimension should be clear from the context. For example the unity matrix may be denoted $I$ independent of dimension. A few specific matrices are denoted by lower case greek letters as e.g. the Pauli matrices $\sigma_j$ and the left-(right-) isoclinic generators $\rho_j$ ($\lambda_j$). 

\section{The Jones description}
The transverse electric field components of a plane electromagnetic wave,  propagating in the $z-$direction, has only two components, and can thus be described as the column vector 
\begin{equation}
\bs e=\begin{pmatrix} e_x \\ e_y  \end{pmatrix}.
\end{equation}
We refer to this vector as the complex \emph{Jones vector}.
Often the Jones vector is given in the frequency domain, so we will allow it to be frequency dependent, although in this paper this dependence is not important, so we will not write it out explicitly. The complex vector elements describe the $x-$ and $y-$ components of the electric field, and their phase relative to some phase reference. For most purposes, e.g. in coherent optical signaling,  it suffices to describe how the Jones vector is transformed when propagating in the system.  A linear transmission medium can then be described by its complex transfer function, and for electromagnetic waves this generalizes to a complex $2 \times 2$ Jones matrix $T$, that relates the input and output Jones vectors via 
\begin{equation}
\bs e(z)= T(z)   \bs e(0)
\label{eq:jonestrans}
\end{equation}
where $\bs e(0)$ denotes the input $(z=0)$ Jones vector to the system and $z$ is an arbitrary output position. The transfer matrix $T$ can be an arbitrary $2 \times 2$ matrix with complex coefficients, but in this analysis we will restrict ourselves to systems without any polarization-dependent loss (such as transmission fibers) where the signal power $P=\bs e^{*t} \bs e=\bs e^{+} \bs e$ must be conserved. This leads to $T^{-1}= T^+$ so that $T$ will be a \emph{unitary} transformation matrix, belonging to the \emph{unitary} group $U(2)$. The subgroup of unitary $2 \times 2$ matrices with determinant=+1 is called the \emph{special unitary group}, or $SU(2)$. 


Since our purpose is to relate the transfer matrix between different descriptions of the signal vector, we need to parametrize the unitary transformation, and to do that will find it useful to describe the evolution of the signal within the medium by the differential equation
\begin{equation}
i\frac{d \bs e}{dz}=  H \bs e 
\label{evoleqJ}
\end{equation}
where $H$ is a matrix describing the transmission medium properties. Again, for lossless transmission we must have $d(\bs e^+ \bs e)/dz=\bs e^+(H^+-H) \bs e=0$ which implies that the matrix $H=H^+$ is Hermitian (i.e. equal to its conjugate transpose). A useful parameterization of Hermitian matrices is via the Pauli spin matrices $\sigma_j$ (see definition, notation and general properties in appendix \ref{paulimat}). This means that $H$ can be written as a linear combination of the four Pauli spin matrices:
\begin{equation}
H = h_0 \sigma_0+ h_1 \sigma_1+h_2 \sigma_2+ h_3 \sigma_3=h_0 I+\vec h \cdot \vec{\sigma}
\label{matrixh}
\end{equation}
or explicitly
\begin{equation}
H = \left ( \begin{array}{cc} h_{0}+h_1 & h_2-i h_3\\ h_2+ih_3&h_0-h_1 \end{array} \right).
\end{equation}
Here $h_k$ are 4 real coefficients describing the medium, which may be formed into a scalar $h_0$ and a real 3-vector $\vec h=(h_1,h_2,h_3)$. 
Assuming $H$ to be constant enables the differential equation (\ref{evoleqJ}) to be solved using the matrix exponential as
\begin{equation}
\bs e(z)= \exp[ -i H z] \bs e(0)= T \bs e(0).
\label{jdesc}
\end{equation}
The matrix exponential can be expanded in a Taylor series to prove that 
\begin{equation}
T= \exp[ -i H z]= \exp(-i h_0z) U(z \vec h ) \label{jmat}
\end{equation}
where 
\begin{equation}
U(\vec \alpha)=\exp(-i \vec \alpha \cdot \vec \sigma)=[I \cos(\alpha)-i \frac{\vec \alpha}{\alpha}\cdot \vec \sigma \sin(\alpha)]
\label{unitary}
\end{equation}
denotes the generic special unitary transformation, or $SU(2)$ group member, parameterized by the real 3-component vector $\vec \alpha$, the modulus of which is the scalar $\alpha=|\vec \alpha|$. Physically the parameterization of the medium in $h_0$ and $\vec h$ can be identified as follows: The $h_0$ coefficient in the Jones matrix (\ref{jmat}) corresponds to a common phase change of both field components, and the remaining coefficients define a unitary transformation $U$ via (\ref{unitary}) that gives a polarization state change. The simplest non-trivial examples are (i) $\vec h=(h_1,0,0)$, for which $T$ models a linear retarder plate with phase retardation $\pm h_1 z$ in the $x-$ and $y-$ polarizations, (ii) $\hat{h}=(0,h_2,0)$ which is a retardation $\pm h_2 z$ between the linear polarization states at $\pm 45$ degrees relative to $x$ and $y$, and (iii) $\hat{h}=(0,0,h_3)$ which is a rotator (circular retarder) that rotates x and y an angle $h_3 z$.  In the next sections we will describe this polarization transfer matrix in the Stokes-Mueller and 4d signal spaces, respectively.

\section{The Stokes-Mueller description}

The 4-component Stokes vector corresponding to the Jones vector $\bs e$ has real components given by the scalar $P=\bs e^+  \sigma_0 \bs e=\bs e^+  \bs e$ (which is simply the optical power), and the real 3-vector $\vec e=\bs e^+ \vec \sigma \bs e$. However, for fully polarized light and absence of polarization-dependent losses, it is sufficient to consider the evolution of the vector $\vec e$ only. As a consequence the transformation (Mueller) matrices will be $3 \times 3$. Explicitly, we can express $\vec e$ in the complex field components as
\begin{equation}
\vec e=\left [ |e_x|^2-|e_y|^2, 2 \re (e_x e_y^*), -2 \im(e_x e_y^*) \right ]^t.
\label{stokesdef}
\end{equation}
Note that any common or \emph{absolute} phase of $e_x$ and $e_y$ will not be modeled by the Stokes vector $\vec e$.
Since only the direction of the Stokes vector $\vec e$ will change it is often described as a point on a sphere, called the Poincar\'e sphere. 
We may then write the evolution equation for $\vec e$ by using the definitions and (\ref{evoleqJ}-\ref{matrixh}) to obtain
\begin{equation}
\frac{d \vec e}{dz}= \bs e^+ i(H^+ \vec \sigma - \vec \sigma H)  \bs e
=2\vec h \times \vec e
\label{evoleqS}
\end{equation}
where  $\vec h \times$ denotes the cross product operator
\begin{equation}
\vec h \times =  \left ( \begin{array}{ccc} 0 & -h_3 &h_2 \\ h_3 & 0 & -h_1\\ -h_2 & h_1 & 0 \end{array} \right).
\label{crossprodop}
\end{equation}
The evolution equation (\ref{evoleqS}) has the well-known geometrical interpretation of describing a vector $\vec e$ rotating around an axis directed along $\vec h$.  In the case of constant coefficients this equation can be integrated, again using the matrix exponential, to
\begin{equation}
\vec e(z)= \exp(z2\vec h \times) \vec  e(0)=  M (z2\vec h) \vec e(0)
\label{eq:smtrans}
\end{equation}
where $M$ denotes the Mueller matrix, which relates the input and output Stokes vectors. In this case when we have no polarization-dependent losses, the Mueller matrix is a rotation matrix, generally expressed as
\begin{gather}
M(\vec \alpha)= \exp[ \vec \alpha \times]=  \nonumber \\
I  + \sin(\alpha) \frac{\vec \alpha \times}{\alpha}   +\frac{1-\cos(\alpha)}{\alpha^2} (\vec \alpha \times)^2= \nonumber \\ 
I \cos(\alpha) + \sin(\alpha)( \hat{\alpha} \times)  + (1-\cos( \alpha)) (\hat{\alpha}\hat{\alpha} \cdot) \label{orth}
\end{gather}
which describes a rotation an angle $\alpha=| \vec \alpha |$ around the unit vector $\hat \alpha=\vec \alpha/ \alpha$. 
Therefore the Mueller matrix $M$ is often described geometrically with the Poincar\'e sphere rotating around the birefringence vector $\vec h$. The set of real $n \times n$ matrices $M$ preserving a real vector length must fulfill $M^{-1}=M^t$, and forms the \emph{orthogonal}  group $O_3$. The subset of $O_3$ with determinant=+1 forms the $O_3^+$ group.

By comparing the Jones transfer matrix (\ref{unitary}) with the Mueller matrix (\ref{orth}) we have a mapping between the two formalisms that illustrates the isomorphism between the $O_3^+$ matrices $R(\vec \alpha)$ and the $SU(2)$ matrices $U(\vec \alpha)$. However the Jones description (although not the $SU(2)$ group) is more complete in that it may also contain the absolute phase change on the wave from the medium (the $h_0$ coefficient in (\ref{unitary}), which is absent from the Stokes-Mueller approach. On the other hand, the Stokes-Mueller approach facilitates the geometrically pleasing, and also intuitive, interpretation of rotating polarization vectors on the Poincar\'e sphere. 

Finally we should mention the derivation made in the end of Appendix \ref{paulimat} which connects the coherency matrix $\bs e \bs e^+$, see (\ref{coherencymat}), and the corresponding Stokes vector $\vec e$ via

\begin{equation}
 \bm e \bm e^+=\frac{P+\vec e\cdot \vec \sigma}{2}.\label{2dcoherency}
\end{equation}
If we have two different Jones vectors $\bs e_{1,2}$ with corresponding Stokes vectors $\vec e_{1,2}$ and powers $P_{1,2}=|\vec e_{1,2}|$ we can use this relation to obtain the useful scalar product (or projection) formula
\begin{equation}
| \bm e_1^+ \bm e_2|^2=\frac{P_1 P_2+\vec e_1\cdot \vec e_2}{2}. \label{jonesprojection}
\end{equation}
connecting Jones and Stokes state vectors.

So far we have done nothing new, but the preceding relations between the Jones and Stokes-Mueller descriptions are well known, and can be found in numerous books and articles on the topic, e.g. \cite{damask05,frigo86,gordon00}. We provide them for reference and to enable easy connections with a third and less known description, based on the four-dimensional Euclidean space that will be introduced in the next section.


\section{Four-dimensional signal space description} \label{sec:4dsig}
Communication theory analyze signals in a signaling space that is an $n$-dimensional vector space containing the set of transmitted signals (the signaling \emph{constellation}), as a set of discrete points in $n$-d. The received signals belong to this set, possibly perturbed by distortions, rotations, and noise. The geometric description of signals in signal space is valuable, because the Euclidean distances between the constellation points will directly translate into robustness to additive noise \cite{agrell09}. For coherent optical communications the signaling space is four-dimensional (4d), so a real 4d description of the electric field is in many cases preferred over the Jones or Stokes-Mueller formalisms in this context \cite{agrell09,betti91,cusani92}. 

The Jones description can be taken to an equivalent 4d description by expressing the complex Jones vector $\bs e$ as the real 4-vector 
\begin{equation}
\vec E(z)=\left( \begin{array}{c} \re(e_x)\\ \im(e_x) \\ \re(e_y) \\ \im(e_y) \end{array}\right).
\end{equation}
A Jones transformation such as (\ref{jdesc}) can now be written as
\begin{equation}
\vec E(z)=  N(z)  \vec E(0)
\label{eq:4d}
\end{equation}
where $N(z)$ is a real 4d transformation matrix. As we noted for the Jones and Stokes descriptions, the absence of polarization-dependent loss or gain restricts the power $P=\vec E^t \vec E$ to be conserved and thus $N^T N=I$ so the $N$ is a 4d orthogonal matrix, belonging to the $O_4$ group. In order to identify the elements of $N$ with those of the Jones matrix $T$ in (\ref{jdesc}) we study the evolution equation for the vector $\vec E$, i.e.
\begin{equation}
\frac{d \vec E }{dz}= -K \vec E
\label{evoleq4}
\end{equation}
for some matrix $K$ describing the medium. Since $P$ must be constant with respect to $z$, we can show that $K$ should satisfy $K^t=-K$, i.e. $K$ should be \emph{skew-symmetric}. This means that its four diagonal elements vanish, and only the 12 off-diagonal elements of $K$ remain. By separating (\ref{evoleqJ}) in real and imaginary parts, and using (\ref{matrixh}) we find that the matrix $ K$ equals
\begin{gather}
K= \left ( \begin{array}{cccc} 0 & -(h_0+h_1) & h_3 & -h_2\\ h_0+h_1 & 0 & h_2 &h_3 \\ -h_3 & -h_2 & 0 & -(h_0-h_1)\\ h_2 & -h_3 & h_0-h_1 & 0 \end{array} \right) = \nonumber \\
(-i)(h_0 D_{03}+ h_1 D_{13}+ h_2D_{23})+i h_3 D_{30} . \label{eq:kmat}
\end{gather}
Here $D_{jk}$ with $j,k \in \{0,1,2,3\}$ denote the 16 \emph{Dirac matrices}. The Dirac matrices are $4 \times 4$ matrices with complex coefficients that are a 4d generalization of the Pauli spin matrices, being defined via the Kronecker product as $D_{jk}= \sigma_j \otimes \sigma_k$. In appendix \ref{diracmatrices} they are tabulated and their basic properties described. In analogy with the previous sections we can integrate (\ref{evoleq4}) if the $h_k$ coefficients are constant, and describe the 4d rotation using the matrix exponential as 
\begin{equation}
\vec E(z)= \exp(-zK) \vec  E(0).  \label{eq:4dtrans}
\end{equation}
This description can be further simplified, but for that we need to describe the general properties of 4d rotation matrices, which will be done in the next section.

It should be mentioned that the Dirac matrices have been used previously in connection with polarization optics. The recent paper by Tahir et al. \cite{tahir10}, used the definition (\ref{diracdef}) of the Dirac matrices to express the coherency matrices. However the 4d rotation description for polarization transformation that we pursue here was not discussed. Nontheless, that connection was indeed made in another recent paper, by Baumgarten \cite{baumgarten11}. That work applied a set of real Dirac matrices (obtained by replacing $\sigma_3$ with $i \sigma_3$ in our definition (\ref{diracdef}), and pioneered by Majorana in the 1930:ies \cite{wilczek09}), both to describe relativistic electromagnetics, as well as to connect them to (3d) rotation matrices and Lorentz transformations. Baumgarten's work could thus probably help with a future extension of the present work to the non-unitary case with polarization-dependent losses, where the Mueller matrices belong to the Lorentz group.

\section{Parametrization of 4d rotations}
The argument in the preceding section, that a rotation should preserve the vector length, can be generalized to show that a rotation matrix $ R=\exp(L)$ in $n$-d can be expressed as the exponential of a skew-symmetric matrix  $n\times n$ matrix $L$. Such a matrix has elements $L_{ij}=-L_{ji}$ so that the diagonal elements must be zero, and the $n(n-1)/2$ elements above the diagonal is enough to describe all degrees of freedom. In the special case of 4d we have 6 degrees of freedom, i.e. we need 6 parameters to describe an arbitrary 4d rotation. It is useful to use a subset of the Dirac matrices to describe $L$, and the 6 skew-symmetric Dirac matrices are the $D_{ij}$ where one (but not both) of $i$ and $j$ equals 3. Note that these are the Dirac matrices with purely imaginary elements, so they are multiplied with the $i$ in the below formulas to be made real. In appendix \ref{4drot} we present a complementary discussion of 4d rotations and their nomenclature. 

We will thus introduce the following 6 matrices as the 4d rotation basis  (notation will be motivated later):
\begin{eqnarray}
\rho_1=  (-i) D_{13}= \left(
\begin{array}{cccc}
  0 &-1 & 0 & 0    \\
  1 & 0 & 0 & 0 \\
  0 & 0 & 0 & 1 \\
  0 & 0 & -1 & 0
\end{array}
\right) \label{rho1}\\
\rho_2= (-i) D_{23}= \left(
\begin{array}{cccc}
  0 &0 & 0 & -1    \\
  0 & 0 & 1 & 0 \\
  0 & -1 & 0 & 0 \\
  1 & 0 & 0 & 0
\end{array} 
\right) \\
\rho_3= (i) D_{30} =\left(
\begin{array}{cccc}
  0 &0 & 1& 0    \\
  0 & 0 & 0 & 1 \\
  -1 & 0 & 0 & 0 \\
  0 & -1 & 0 & 0
\end{array}
\right) \label{rho3}
\end{eqnarray}
which forms the vector $\vec{\rho}=(\rho_1,\rho_2, \rho_3)$ and 
\begin{eqnarray}
\lambda _1= (i)D_{03}=\left(
\begin{array}{cccc}
  0 &1 & 0 & 0    \\
  -1 & 0 & 0 & 0 \\
  0 & 0 & 0 & 1 \\
  0 & 0 & -1 & 0
\end{array}
\right) \\
\lambda_2=(-i) D_{32} =\left(
\begin{array}{cccc}
  0 &0 & 0 & -1    \\
  0 & 0 & -1 & 0 \\
  0 & 1 & 0 & 0 \\
  1 & 0 & 0 & 0
\end{array}
\right) \\
\lambda_3=(i)D_{31}= \left(
\begin{array}{cccc}
  0 &0 & 1& 0    \\
  0 & 0 & 0 & -1 \\
  -1 & 0 & 0 & 0 \\
  0 & 1 & 0 & 0
\end{array}
\right) \label{lambda3}
\end{eqnarray}
which forms the vector $\vec\lambda=(\lambda_1,\lambda_2,\lambda_3)$.
An arbitrary real skew-symmetric 4d matrix $L$ can thus be written as a linear combination of these matrices, and an arbitrary rotation
\begin{equation}
R(\vec \alpha,\vec \beta)=\exp(L)=\exp(-\vec \alpha \cdot \vec \rho-\vec \beta \cdot \vec \lambda),\label{eq:R}
 \end{equation}
 i.e. as a linear combination of the six above matrices where $\vec \alpha, \vec \beta$ are two arbitrary, real, three-vectors parameterizing the rotation. The negative sign is chosen to conform with the previous sections. We can thus call the six matrices $\vec \rho, \vec \lambda$ the \emph{generating matrices} for 4d rotations. We may also observe that the following relations hold:
\begin{eqnarray}
\rho_k^2=\lambda_k^2=-I  \hspace{2mm} \forall k \nonumber \\
\rho_1 \rho_2 \rho_3=-\rho_2 \rho_1 \rho_3 = I  \nonumber \\
\lambda_1 \lambda_2 \lambda_3=-\lambda_2 \lambda_1 \lambda_3=I, \label{rholambda}
\end{eqnarray}
where $ I$ is the 4d unity matrix. These multiplication rules can be used to show that matrices formed by a linear combination of the $\rho_k$ and the unity matrix form a multiplicative \emph{subgroup} to the group of all real 4d skew-symmetric matrices. The same holds for the group of matrices formed by a linear combination of $\lambda_k$ and the unity matrix.
Quite surprisingly these two subgroubs \emph{commute}, since
\begin{equation}
\rho_k \lambda_j=\lambda_j \rho_k \label{rotl}
\end{equation}
for all $j$ and $k$. Note that multiplications within each subgroup is non-commuting, according to (\ref{rholambda}). This has important consequences for the rotations; since we may, for commuting operators, decompose the matrix exponential to the matrix product
\begin{equation}
R=R_R R_L=R_L R_R
\label{eq:r}
\end{equation}
where
\begin{gather}
R_R(\vec \alpha)=\exp(-\vec \alpha \cdot \vec \rho) \\
R_L(\vec \beta)=\exp(-\vec \beta \cdot \vec \lambda) .
\end{gather}
This implies that we can decompose an arbitrary 4d rotation in to two commuting subgroups, which are often denoted \emph{right-isoclinic}, $R_R$ and \emph{left-isoclinic}, $R_{L}$, rotations respectively\cite{4drot}. Thus any 4d rotation can be written as one right-isoclinic rotation followed by a left-isoclinic rotation, or vice versa. 

The isoclinic rotations can be written in a more explicit form by evaluating the matrix exponential and using 
\begin{equation}(\vec \alpha \cdot \vec \rho)^2=-I \vec \alpha^2
\end{equation}
to obtain
\begin {gather}
R_R=\exp[-\vec \alpha \cdot \vec  \rho]=\exp[-\alpha \hat \alpha \cdot \vec  \rho]= \nonumber \\
\sum_{k=0}^\infty \frac{(-\alpha \hat \alpha  \cdot  \vec \rho)^k}{k !}= I - (\hat \alpha \cdot  \vec \rho) \alpha - I \frac{\alpha^2}{2}+ (\hat \alpha \cdot  \vec \rho) \frac{\alpha^3}{6}+... \nonumber \\
= I \cos(\alpha) - (\hat \alpha \cdot  \vec \rho) \sin(\alpha). \label{eq:rl}
\end{gather}
Analogously, we find for the left-isoclinic rotations
\begin {equation}
R_L= I \cos(\beta) - (\hat \beta \cdot  \vec \lambda) \sin(\beta). \label{eq:rr}
\end{equation}
Eqs. (\ref{eq:r},\ref{eq:rl},\ref{eq:rr}) give us a full, 6 DOF parameterization of the 4d rotations. 

We can now return to Sec. \ref{sec:4dsig} and identify the elements $h_0, \vec h$ from the matrix $K$ in (\ref{eq:kmat}) with the general 4d exponent $L$. This gives $\vec \alpha=\vec h$ and  $\vec \beta=(h_0,0,0)$. We can thus express the 4d signal (\ref{eq:4dtrans}) in terms of 4d rotations as
\begin {gather}
\vec E(z)= \exp(-z \vec h \cdot \vec \rho) \exp(-z (h_0,0,0)\cdot \vec \lambda) \vec  E(0).  \label{eq:4dtrans1}
\end{gather}
\begin{table*}[t!]
\caption{Relation between the conventional Jones, Stokes-Mueller and 4d vector descriptions of electromagnetic transformations. The transformation is characterized by four parameters; an absolute phase shift $\alpha_0$ and a polarization change described by the real 3-vector $\vec \alpha=\alpha \hat \alpha$. The vectors $\vec \sigma, \vec \rho, \vec \lambda$ are three vectors formed by the Pauli spin matrices, and the left- and right-isoclinic 4d rotation generators, respectively, which are defined in Eqs. (\ref{rho1}-\ref{lambda3}). Note that the absolute phase shift is absent from the Stokes-Mueller description.\label{classicaltable} }
\begin{center}
\begin{tabular}{|c|c|c|c|c|} 
\hline
& Signal Vector & Input/Output Transformation & Transfer Matrix  \\ \hline \hline
Jones & $\bs e=\left( \begin{array}{c} e_x \\ e_y  \end{array} \right) $ &  $\bs e_\text{out}=T \bs e_\text{in}$& $\begin{aligned} T= &  \exp[-i \alpha_0-i\vec  \alpha \cdot \vec \sigma] =\\Ê
& \exp(-i \alpha_0) [I \cos(\alpha)-i \hat \alpha \cdot \vec \sigma \sin(\alpha)] \end{aligned} $\\ \hline
Stokes-Mueller & $\vec e=\begin{pmatrix} |e_x|^2-|e_y|^2 \\ 2 \re (e_x e_y^*) \\ -2 \im(e_x e_y^*) \end{pmatrix} $ & $\vec e_\text{out}=M \vec e_\text{in} $ & $ \begin{aligned} M= & \exp[2 \vec \alpha \times ]= \\
& I \cos(2\alpha) + \sin(2\alpha) \hat{\alpha} \times  + (1-\cos( 2\alpha)) \hat{\alpha}(\hat{\alpha} \cdot) \end{aligned} $\\ \hline
Four-dimensional& $\vec E=\begin{pmatrix} \re(e_x)\\ \im(e_x) \\ \re(e_y) \\ \im(e_y) \end{pmatrix} $ &$\vec E_\text{out}=R \vec E_\text{in}$ 
& $\begin{aligned} R= & \exp[- (\alpha_0,0,0) \cdot \vec \lambda-\vec  \alpha \cdot \vec \rho] = \\Ê
& \exp(- \alpha_0 \lambda_1) [I \cos(\alpha)-\hat \alpha \cdot \vec \rho \sin(\alpha)] \end{aligned} $\\
\hline
\end{tabular}
\end{center}
\end{table*}%

This completes the description of polarization transformations in terms of 4d rotations, and the result is that we now have three alternative descriptions of polarization transformations. These are the complex Jones calculus in Eq. (\ref{eq:jonestrans}), the real Stokes-Mueller description in Eq. (\ref{eq:smtrans}) and the real 4d rotation description from Eq. (\ref{eq:4dtrans1}). Those have, for convenience been summarized in Table \ref{classicaltable}.

The above concludes a relatively straightforward extension of the Jones calculus to 4d, that in itself might be useful in e.g. fiber communication analysis\cite{Karlsson10ecoc}, but it provides nothing really physically novel. However, the remainder of this paper will be devoted to the 4d rotations described by the left-isoclinic matrices $\lambda_2$ and $\lambda_3$, which seem to have no correspondence in the classical Jones- or Stokes-vector descriptions. As we saw above, only the right-isoclinic rotations (\ref{eq:rr}) of the 4d $E$-field vector have a physical realization in terms of the corresponding Jones  and Mueller matrices. Obviously the $\lambda_1$-component gives a constant phase shift of the field vector, so that has a clear physical interpretation. However, the $\lambda_2-$ and $\lambda_3-$components describe transformations that are not used within the classical Jones- or Stokes-Mueller calculus, and the reason is simply that they cannot be realized as physical transformations of the electromagnetic field. In fact, a propagating electromagnetic wave cannot undergo such a change, irrespective of the physical realization. The proof of this statement lies in that those state changes will not satisfy the bosonic commutation relations, as will be motivated in next section and appendix \ref{forbidden}. We will therefore refer to those transformations as \emph{forbidden rotations}. We should emphasize that the existence of the forbidden rotations is not a shortcoming of the known polarization calculi. It is merely a richer property of the 4d real space that is not needed for the description of propagating waves, but which can be taken advantage of if we understand it. For example if we can take the absolute phase rotation (which in terms of 4d rotations is given by the $\exp(i \varphi \lambda_1)$-matrix) to Stokes space, we may find a natural extension of the absolute phase rotations to Stokes space, a problem that researchers have worked on previously \cite{Frigo09}. 

A first clue to an understanding of the forbidden rotations will be to go back to the Jones- and Stokes-Mueller formalisms, and attempt to express these transformations there. We will start by describing how the Jones-vector formalism can be extended.

\section{The forbidden rotations in Jones space}
The forbidden (left-isoclinic) transformation acting on the 4d vector $\vec A=\begin{pmatrix} a_x \\a_y \end{pmatrix}$ is
\begin {gather}
\vec A_\text{out}= 
\exp(-\beta \hat \beta \cdot \vec \lambda) \vec  A_\text{in}
\end{gather}
where subscript denote input/output states. If we write this in complex form, by collecting the proper real and imaginary parts, it can be expressed as
\begin{gather}
\begin{pmatrix} a_x \\ a_y^*\end{pmatrix}_\text{out} = 
U(\vec \beta)
\begin{pmatrix} a_x \\ a_y^*\end{pmatrix}_\text{in} \label{eq:ri1}
\end{gather}
%
where 
$U$ is the unitary transformation defined in (\ref{unitary}). Thus, we obtain the a unitary transformation as in section 2, but now acting on the vector $[a_x, a_y^*]^t$, i.e. the Jones vector with the y-component conjugated. Thus, to describe the left-isoclinic transformations with Jones calculus, we \emph{need to redefine the Jones state vector}. Such transformation vectors arise e.g. in the context of parametric amplifiers \cite{louisell61}, but then the transfer matrix is not unitary. In fact, in appendix D we show that such a transformation is only consistent with the quantum mechanical formulation of polarization operators if the two components $\beta_2=\beta_3=0$, for which (\ref{eq:ri1}) reduces to
\begin{equation}
\begin{pmatrix} a_x \\ a_y^*\end{pmatrix}_\text{out} = 
\begin{pmatrix} \label{eq:ri2}
\exp(i \beta_1) &  0 \\
0 & \exp(-i \beta_1 )
\end{pmatrix}
\begin{pmatrix} a_x \\ a_y^*\end{pmatrix}_\text{in}
\end{equation}
 which is the same phase rotation applied to both field components. Therefore changes described by the $\lambda_{2}$ and $\lambda_3$ rotation generators are indeed forbidden for the electromagnetic field. However this does not preclude a mathematical description of them. In the following discussion we will describe the right (left-) isoclinic rotations with the dimensionless parameters $\vec \alpha$ ($\vec \beta$). 
 
We may unify the left- and right-isoclinic transformations in an elegant way by replacing the Jones vector with a \emph{Jones state matrix}
\begin{equation}
\bs E=\begin{pmatrix}
e_x &  e_y^* \\
e_y & -e_x^*
\end{pmatrix} \label{jonesstate}.
\end{equation}
With this definition a right-isoclinic (i.e. the conventional) transformation is performed by applying a unitary transformation to the two column vectors. Similarly, a left-isoclinic (comprising the phase and forbidden) transformation is realized by applying a unitary transformation to the row vectors. The first column vector is the conventional Jones vector and the second is the corresponding orthogonal state of polarization, and obviously these two must transform with the same unitary matrix, so the transformation $U \bs E$ is just a trivial extension of the standard Jones analysis. However, the left-isoclinic rotation are described by a unitary operation on the transposed matrix, i.e. by $U \bs E^t=(\bs E U^t)^t$. Thus we can describe the left-isoclinic transformations by left multiplying the state matrix with a transposed unitary matrix. The full input-output relation for the general 4d rotation becomes
\begin{equation}
\vec E_\text{out}=R(\vec \alpha,\vec \beta) \vec E_\text{in}
\label{gen4drot}
\end{equation}
where $R$ is given by (\ref{eq:R}), can thus be compactly expressed by the Jones state matrix as
\begin{eqnarray}
\bs E_\text{out}=
U(\vec \alpha) \bs E_\text{in} U(\vec \beta)^t \label{genjones}
\end{eqnarray} 
where again $U$ denotes the unitary matrix as defined in (\ref{unitary}). In this way we can recover the full 6-parameter rotation family by left and right multiplying the Jones state matrix with unitary matrices. It is noteworthy that the Jones state matrix is of unitary form, but with determinant $-P$, which means that its unitary structure (\ref{jonesstate}) is preserved by the transformation (\ref{genjones}), as expected. The commutation between right- and left-isoclinic rotations is also evident in  (\ref{genjones}), due to the associativity of matrix products.
\begin{table}[t]
\caption{Overview over the use of the 16 Dirac matrices in terms of the matrices used in the generalized Stokes calculus.}\label{diracmatoverview}
\begin{tabular}{|p{2cm}|p{2cm}|p{2cm}|p{2cm}|} 
\hline
$D_{00}=I$ & $D_{01}=-s_{33}$ & $D_{02}=s_{32} $& $D_{03}=i \lambda_1 $\\ \hline
$D_{10}=s_{11}$ & $D_{11}=-s_{22}$ & $D_{12}=-s_{23} $& $D_{13}=i \rho_1 $\\ \hline
$D_{20}=s_{21} $ & $D_{21}=s_{12} $ & $D_{22}=s_{13}$& $D_{23}= i \rho_2 $ \\ \hline
$D_{30}=-i \rho_3 $ & $D_{31}=-i \lambda_3 $ & $D_{32}=-i\lambda_2 $& $D_{33}=-s_{31} $ \\ 
\hline
\end{tabular}
\end{table}

\section{The generalized Stokes calculus} \label{genstokes}
We saw that the 4d rotations could be described in Jones space by extending the Jones vector to a matrix that is transformed by right- and left multiplication by unitary matrices. The same thing happens when we try to connect the forbidden rotations with the Stokes-Mueller description. We divide this section into two parts; first we present definitions and properties; and second follows a few basic examples.

\subsection{The Stokes state matrix}
In order to describe the forbidden rotations, we need to replace the Stokes vector with a $3 \times 3$ matrix $E$, which we will refer to as the \emph{Stokes state matrix}, defined in terms of the complex field components $e_x$ and $e_y$ as
\begin{equation}
E=\begin{pmatrix} |e_x|^2-|e_y|^2 &2 \re(e_x e_y) & 2 \im(e_x e_y) \\ 2 \re (e_x e_y^*) & -\re(e_x^2-e_y^2) & -\im(e_x^2-e_y^2) \\ -2 \im(e_x e_y^*) & \im (e_x^2+e_y^2) & -\re(e_x^2+e_y^2) \end{pmatrix} \label{smat}.
\end{equation}
By using the 4d vector $\vec E$ and the Dirac matrices $D_{ij}$ defined in Appendix B we can also express the Stokes state matrix as
\begin{equation}
E_{jk}=\vec E^t s_{jk} \vec E
\label{s_to_e}
\end{equation}
where $s_{jk}$ is a $3 \times 3$ matrix with components given by the $4 \times 4$ Dirac matrices as
\begin{equation}
s_{jk}=\begin{pmatrix} D_{10} &D_{21} & D_{22} \\ D_{20} & -D_{11} & -D_{12} \\ -D_{33} & D_{02} & -D_{01} \end{pmatrix}.
\label{smat2}
\end{equation}

With these definitions we can, cf. Appendix \ref{stokesmatrixderivation} for details, express the general 4d rotation (\ref{gen4drot}) as
\begin{equation}
E_\text{out}=M(\vec \alpha) E_\text{in} M(\vec \beta)^t
\label{newstokes}
\end{equation}
where subscripts $_\text{in/out}$ refers to the input and output states. We note that the right- (left-) isoclinic rotations is obtained by right- (left-) multiplication of the Stokes matrix with the corresponding rotation matrix. 
It is an interesting observation that the 6 rotation generators (\ref{rho1})-(\ref{lambda3}) together with the above 9 Stokes matrix components $s_{ij}$ and the unity matrix comprise all 16 Dirac matrices. This dependence is summarized in Table \ref{diracmatoverview}.

The transformation (\ref{newstokes}) shows that the classical (right-isoclinic) transformations acts by rotating the column vectors by the rotation matrix $M(\vec \alpha)$, in full compatibility with the classical Mueller-Stokes description in (\ref{eq:smtrans}-ref{orth}). The novelty is the phase and forbidden (left-isoclinic) transformations, that acts on the row vectors instead via right multiplication of the rotation matrix $M(\vec \beta)^t$. 

The Stokes state matrix $E$ is a very interesting generalization of the Stokes vector $\vec e$.
Some properties worth highlighting are:
\begin{itemize}
\item The first column, $E_{j1}$ equals the conventional Stokes vector $\vec e$, see (\ref{stokesdef}), which does not depend on any common phase of the electromagnetic field components (which we refer to as the absolute phase). 
\item We define the vectors $\vec f$ and $\vec g$ from the remaining two column vectors, $E_{j2}=\vec f$ and $E_{j3}=\vec g$, which  are \emph{alternative Stokes vectors} that, in absence of left-isoclinic transformations ($\vec \beta=0$), transform exactly as the traditional Stokes vector $\vec e=E_{j1}$. The existence of such alternative Stokes vectors is believed to be reported here for the first time. The reason is probably that they involve the absolute phase of the field, that is not easily measurable in experiments.

\item $E$ is an orthogonal matrix, with the three column vectors $[\vec e,\vec f, \vec g]$ (or the three row vectors) mutually orthogonal. The length of each row or column vector equals the power, $P=|e_x|^2+|e_y|^2$. 

\item The triplet $[\vec e,\vec f, \vec g]$ forms a right-handed coordinate system; i.e. the determinant $\det(E/P)=1$. Any right-isoclinic transformation can be seen as a rotation of this coordinate system. 

\item The triplet formed by the row vectors $[E_{1j},E_{2j},E_{3j}]$ is another right-hand system that is transformed by the left-isoclinic transformations.

\item The mapping between (i) the Jones  state matrix (and 4d vectors) to (ii) Stokes state matrix is 2-to-1, in that both the vectors $\vec E$ and $-\vec E$ will be mapped to the same Stokes state matrix. This is obvious from the definition being proportional to $\vec E$ times itself, cf. (\ref{s_to_e}). 

\end{itemize}
We may thus see that this is an appealing extension of the traditional Stokes vector to account for absolute phase changes via the vectors $\vec f$ and $\vec g$. More mathematical properties of the Stokes state matrix, an explicit relation to the Jones state matrix, and matrix-exponential expressions for the Jones and Stokes state matrices are given in appendix \ref{statematrices}.  

We end this subsection by generalizing the projection formula (\ref{jonesprojection}) to a 4d vectors $\vec E$. In appendix \ref{diracmatrices} we derive the relation (\ref{4dcoherency})
\begin{equation}
\vec E^t \vec E= \frac{P+E_{ij} s_{ij}}{4} 
\end{equation}
which can be seen as a generalization of the 2d coherency relation (\ref{2dcoherency}) to 4d by using the Stokes state matrix $E$ and its elements $E_{ij}$. We can use this formula to derive the scalar product relation for two different 4d vectors $\vec E_1$ and $\vec E_2$, as
\begin{equation}
(\vec E_1 \cdot \vec E_2)^2= \frac{P_1P_2+\vec e_1 \cdot \vec e_2+\vec f_1 \cdot \vec f_2+\vec g_1 \cdot \vec g_2 }{4}
\end{equation}
where the last terms are the scalar products of the Stokes vector and the two alternative Stokes vectors for the respective fields indicated by subscripts.

\begin{figure*}[t]
\begin{center}
\includegraphics[scale=0.6]{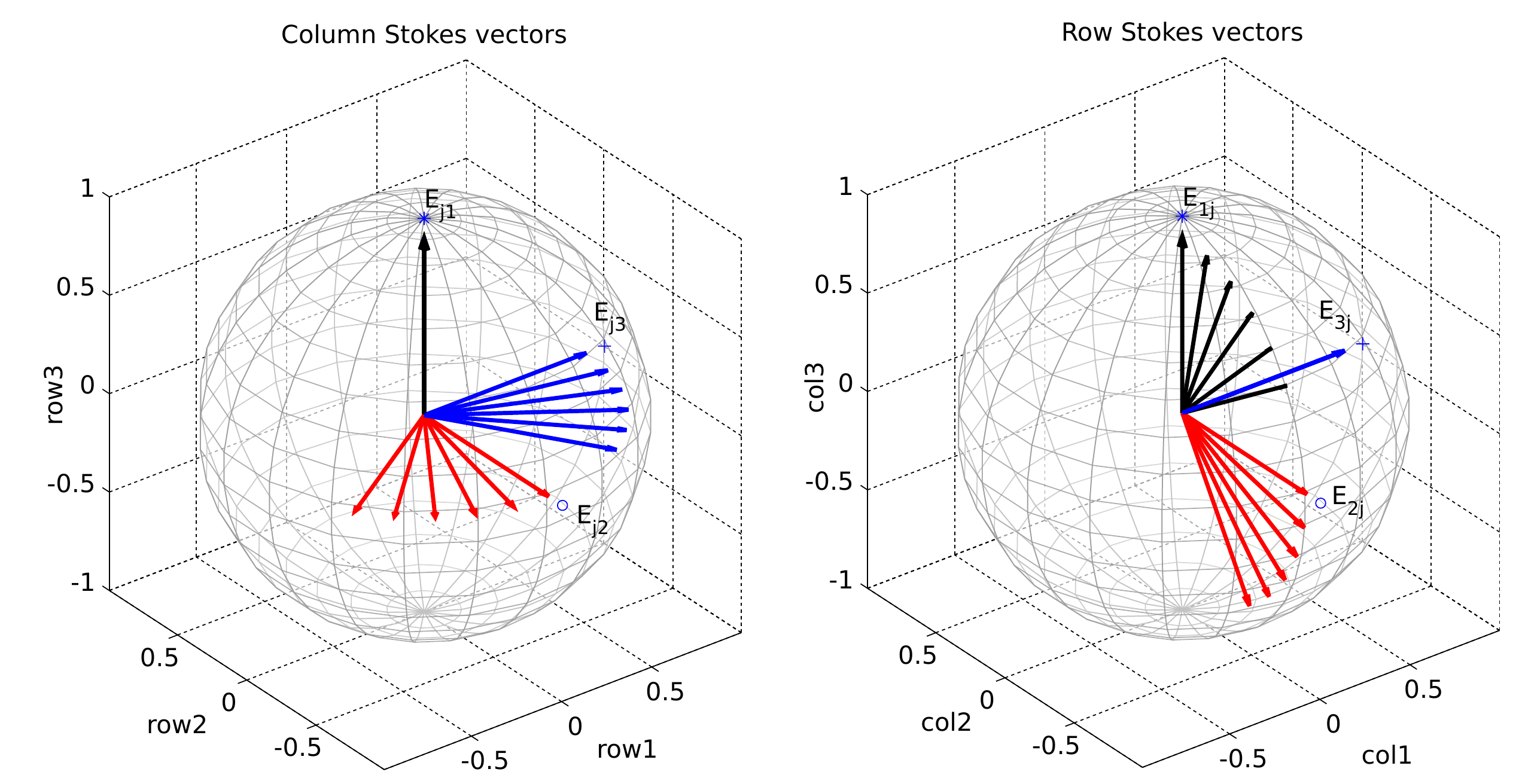}
\caption{Colour) Plot of the column (left) and row (right) vectors of the Stokes matrix for an initial circularly polarized wave with $e_x=1/\sqrt{2}$ and $e_y=i/\sqrt{2}$, affected by a left-isoclinic rotation (\ref{eq:ri2}) with $\vec \beta=(2\phi,0,0)$, i.e. the same constant phase shift $\phi$ in both field components. The phase shift (which in Stokes space gives twice the angular excursion) changes from 0 to 0.5 rad in steps of 0.1. The first, second, and third row/col vectors are colored black, red, and blue, and their initial states are marked with a star, circle and cross, respectively. This kind of polarization transformation which is an absolute phase shift will rotate the column Stokes vectors (left plot) around $E_{j1}$ vector. The row Stokes vectors (right plot), on the other hand will rotate around the first column axis.}
\label{fig:example}
\end{center}
\end{figure*}

\subsection{Using the Stokes state matrix}
A first example is the Stokes state matrix of a linearly polarized wave with common phase $\phi$, i.e. the Jones vector $[ \cos(\theta), \sin(\theta) ]^t \exp(i \phi)$. Its generalized Stokes vectors are 
\begin{align}
\vec e & =[\cos(2 \theta), \sin(2 \theta),0]^t \nonumber \\
\vec f & =[\sin(2 \theta) \cos(2 \phi), -\cos(2 \theta) \cos(2\phi),\sin(2 \phi)]^t \nonumber \\
\vec g & =[\sin(2 \theta) \sin(2 \phi), -\cos(2 \theta) \sin(2\phi),\cos(2 \phi)]^t \nonumber
\end{align}
from which we can see that the absolute phase $\phi$ affects the direction of the vectors $\vec f$ and $\vec g$ in the plane normal to $\vec e$. Thus the direction of $\vec f$ and $\vec g$ is associated with the absolute phase of the state. 

As another example we consider the phase shift of eigenstate propagation, not visible in the classic Stokes-Mueller calculus. Consider the simplest linearly birefringent medium, retarding the $x$ vs. $y$ components a phase $\phi$. The corresponding Jones matrix is
\begin{equation}
T=\begin{pmatrix} \exp(i \phi/2) & 0 \\ 0 & \exp(-i \phi/2) \end{pmatrix}
\end{equation}
showing that for eigenstate (in this case $x$ or $y$-polarized) waves, a pure phase retardation of $\pm \phi$ is incurred, without changing the state of polarization. The corresponding Mueller matrix is (the right isoclininc)
\begin{equation}
M=\begin{pmatrix} 1 & 0 & 0 \\ 0& \cos(\phi) & -\sin(\phi) \\ 0 & \sin(\phi) & \cos(\phi) \end{pmatrix}
\end{equation}
which multiplies the Stokes state matrix $E=[\vec e,\vec f, \vec g]$ from the left. For $x$ ($y$)-polarized light, $\vec e=[1,0,0]^t$ ($\vec e=[-1,0,0]^t$) so that $\vec f$ and $\vec g$ lie in the $s_1=0$-plane. Assuming the absolute phase to be zero (which is no restriction) leads to the $E=\text{diag}(1,-1,-1)$Êfor $x$-polarized light and $E=\text{diag}(-1,1,-1)$ for the $y-$polarized case. When analyzing the movement of $\vec f$ and $\vec g $ for the two cases, we can see that the direction of rotation is opposite, even if the rotation matrix is the same, since $\vec e$ and $\vec f$ have been inverted. 

As a final example on the use of the Stokes state matrix in a transformation system we show a pure (left-isoclinic) phase rotation, ($\vec \alpha=0$, $\vec \beta=(2\phi,0,0)$ ) affecting the initial Stokes state 
\begin{equation}
E_\text{in}=\begin{pmatrix} 0& 0&1\\0 & -1 &0\\ 1 &0 &0 \end{pmatrix}. \nonumber
\end{equation}
which is right-hand circular polarization at zero absolute phase.
After insertion in (\ref{newstokes}) we obtain the output state
\begin{equation}
E_\text{out}=\begin{pmatrix} 0& -\sin(2 \phi) &\cos(2\phi)\\0 & -\cos(2\phi) &-\sin(2 \phi)\\ 1 &0 &0 \end{pmatrix}, \nonumber
\end{equation}
In Fig. \ref{fig:example}, left, we plot the Stokes state matrix column vectors $[E_{j1}, E_{j2}, E_{j3}]=[\vec e,\vec f, \vec g]$ and in Fig \ref{fig:example}, right, we plot the row vectors $[E_{1j}, E_{2j}, E_{3j}]$, as the phase $\phi$ changes from 0 to 0.5 rad in steps of 0.1. In other words we plot the Stokes state matrix for the Jones vector $[\exp(i \phi),i \exp(i \phi)]^t/\sqrt{2}$. In Fig. (\ref{fig:example}, left) we can see the standard Stokes vector (black) $\vec e$ remaining fixed, as it is not affected by absolute phase rotations, whereas the $\vec f$ and $\vec g$ vectors rotate around it. The row vectors on the other hand rotate around the 1-axis, as can be expected from the rotation operator $M((2 \phi,0,0))$.

Thus we can conclude the following fact from the above examples:
An absolute rotation $\phi$ of the optical wave leads to a rotation of $\vec f$Êand $\vec g$ an angle $2 \phi$ around $\vec e$. 
Without the alternative Stokes vectors this rotation would not be easily visualized within in the Stoke-Mueller calculus, although attempts have been made in that direction previously \cite{Frigo09,Pancharatnam56a}.  

\section{The parallel transport phenomena}
We will now, as an example of the use of the Stokes state matrix, describe the \emph{Pancharatnam phase} effect. Since it is often mentioned (and sometimes mistaken for) another phenomenon of similar origin, the \emph{Berry phase}, we will for completeness compare and contrast with this effect as well.

\begin{figure}[t]
\begin{center}
\includegraphics[scale=0.6]{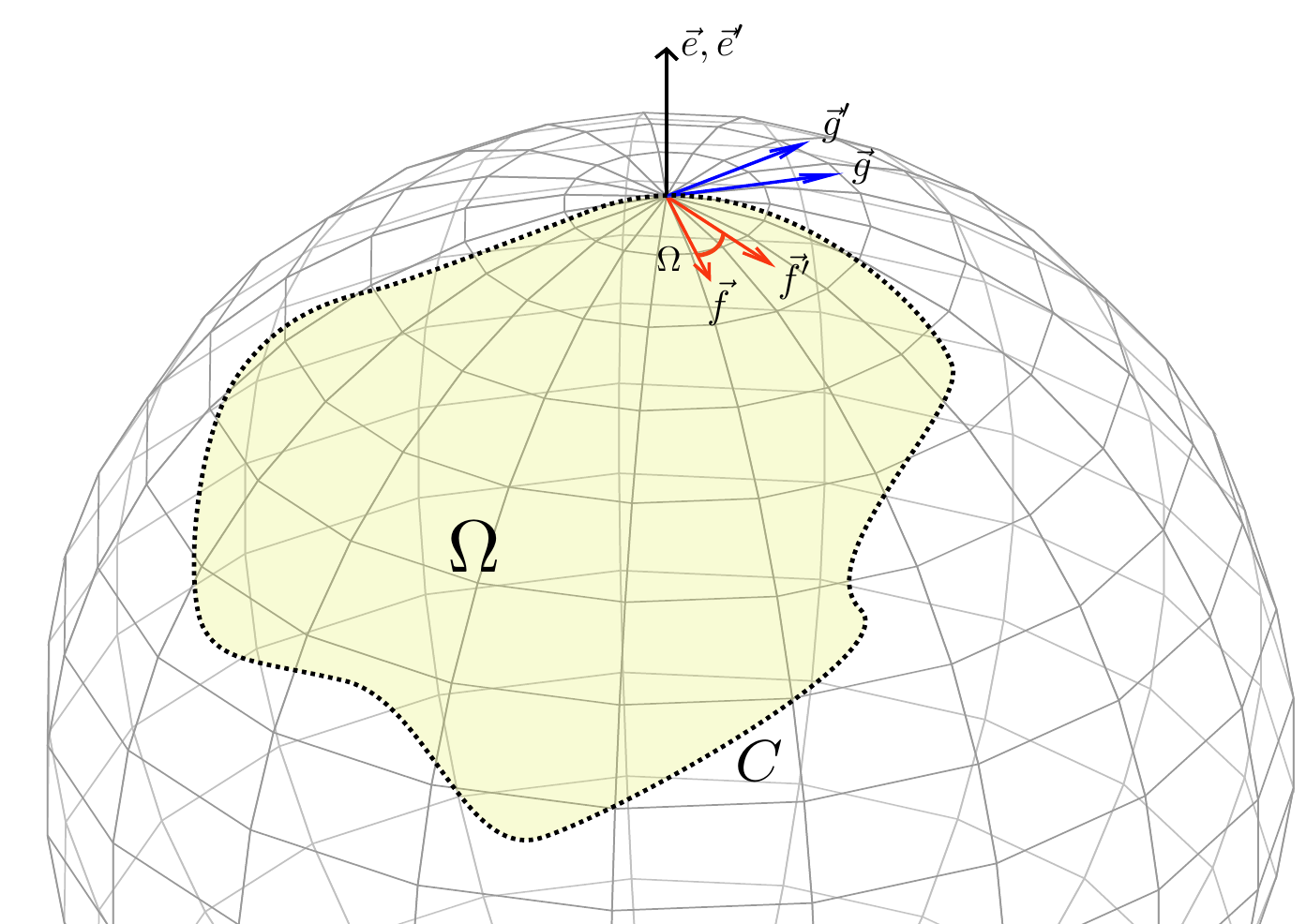}
\caption{Parallel transport of the orthogonal vector triplet $E=[\vec e,\vec f, \vec g]$ where $\vec e$ is radially directed and $\vec f,\vec g$ are tangential to the sphere surface in the starting point (at the north pole). After parallel transport (i.e. translation without rotation) of the triplet around the closed loop $C$, the triplet $[\vec e',\vec f', \vec g']$ where $\vec e=\vec e'$ is formed. The angle between $\vec f$ and $\vec f'$ (and also between $\vec g$ and $\vec g'$) equals $\Omega$, the solid angle subtended by $C$.}
\label{fig:partran}
\end{center}
\end{figure}

Both effects can be intuitively described by the following well-known geometrical fact: Move a vector $\vec f$, parallel to the surface of a sphere, around a closed loop $C$ on the sphere and back to the starting point, and denote that vector $\vec f'$. The movement should be so called \emph{parallel transport}, i.e., a translation without rotation relative to the local coordinate system set up by the coordinate vector, its velocity vector and their cross product. Then $\vec f$ and $\vec f'$ will make an angle $\Omega$, which equals to the solid angle (or unit sphere area) that $C$ encloses. On a Euclidean 2d (non-curved) surface $\vec f$ and $\vec f'$ will obviously coincide, but on a curved surface like that of a sphere they will deviate. This is illustrated in Fig \ref{fig:partran}. 
This phenomenon of \emph{parallel transport} is often used to define curvature, e.g. in the context of general relativity and the study of  metric spaces. 
We will now discuss the two effects in more detail.

\subsection{The Pancharatnam phase}
 In 1956 Pancharatnam \cite{Pancharatnam56a,Pancharatnam56b} wrote two groundbreaking papers with the main purpose of analyzing interference of light with different polarization states. The first paper \cite{Pancharatnam56a} dealt with coherent light, and extends the the Fresnel-Arago interference laws to the case of optical waves with arbitrary states of polarization. The main result was that for interfering a polarization state $S_1$ with another state $S_2$ one must decompose state $S_2$ into one part parallel with state $S_1$ and one part parallel with the orthogonal polarization $-S_1$. This projection involves a phase shift that can be expressed in terms of the area of a spherical triangle formed by the involved polarization states and their orthogonal counterparts on the Poincar\'e sphere. An important consequence of Pancharatnam's analysis, pointed out by Berry \cite{berry87b}, can be stated as follows: If a polarized wave with Stokes vector $S$ evolves through a system so that its polarization state traces out a closed path $C$ with solid angle $\Omega$ on the Poincar\'e sphere, then the starting and finishing states will have an absolute phase difference equal to $\Omega/2$.
 
Instead of using Pancharatnam's spherical trigonometry, this can now be readily explained by using the orthogonal vector triplet $[\vec e,\vec f, \vec g]$, and the phenomenon of parallel transport on a sphere. The vector $\vec e$ is the conventional Stokes vector, orthogonal to the Poincar\'e sphere surface, and it will trace out $C$. The two alternative Stokes vectors $\vec f$ and $\vec g$ will be parallel transported around $C$, and according to the parallel transport phenomenon make an angle $\Omega$ with their initial counterparts. As shown in the previous section, the angular difference between the $\vec f$-vectors of copolarized waves will be half the absolute phase difference, i.e. $\Omega/2$. In this way the Stokes state triplet $[\vec e, \vec f, \vec g] $ becomes a useful tool to describe the Pancharatnam phase change. 

\subsection{The Berry phase}
The Berry phase \cite{berry84} is a very general concept, arising in a number of different physical settings where systems evolve one period while changing a characteristic phase adiabatically. The Aharanov-Bohm effect and Foucault's pendulum are but two manifestations \cite{wilczek89}. In our context of the polarization of light it refers to the polarization change for light propagating along a curved three-dimensional path \cite{Berry87a}. It can be explained as follows. Consider planar polarized light propagating into a system with direction (wave vector) $\vec k$ taking it through an arbitrary path in three dimensions (e.g. a coiled optical fiber), and then leaving the system in the same direction $\vec k$. During its propagation, $\hat  k$ traces out a closed path $C$ on the unit sphere, and the solid angle $\Omega$ of $C$ will exactly equal the angular change of the polarization plane from input to output \cite{Berry87a}. 

Also this phenomenon can be explained by parallel transport, but this time the orthogonal triplet is $[k \hat z, e_x \hat x,e_y \hat y]$, i.e. the wave vector and the two transverse polarization components. Again the parallel transport phenomenon will act by rotating the polarization components (or more specifically the transverse coordinate system) an angle $\Omega$. Contrary to Pancharatnam's phase, which is an absolute phase change induced by a polarization evolution, the Berry's phase is a polarization change induce by a changes in propagation direction. Although the Berry phase for lightwave propagation was extensively investigated in the late 80:ies, it has a much older history. In fact, this purely geometrical effect was originally discovered by Rytov already in 1938 \cite{rytov38} for general electric $\bs E$ and magnetic $\bs H$ field vectors, and specifically evaluated for polarization states by Vladimirskii three years later \cite{vladimirskii41}. The effect is of major practical importance as relating mechanical perturbations to polarization changes in fibers, and it is one of the most significant sources to polarization drift in fiber communications.
\vspace{-1cm}\\


\begin{table*}[t]
\caption{Extension of the conventional Jones and Stokes-Mueller calculii to account for the full 4d rotations, characterized by the 2 real three-vectors  $\vec \alpha$ and $\vec \beta$. The vectors $\vec \sigma, \vec \rho, \vec \lambda$ are three-vectors formed by the Pauli spin matrices, and the left- and right-isoclinic 4d rotation generators, respectively, which are defined in Eqs. (\ref{rho1}-\ref{lambda3}). The unitary matrix $U$ is defined in (\ref{unitary}) and the orthogonal matrix $M$ in (\ref{orth}). The case shown in table \ref{classicaltable} is recovered as the special case $\vec \beta=(\alpha_0,0,0)$.  }
\label{newtable}
\begin{center}
\begin{tabular}{|c|c|c|c|c|} 
\hline
& Signal State & Transformation & Transfer Matrix  \\ \hline \hline
Jones & $\bs E=\begin{pmatrix} e_x & e_y^*\\ e_y & -e_x^* \end{pmatrix}  $ &  $\bs E_\text{out}=U(\vec \alpha) \bs E_\text{in} U(\vec \beta)^t $& $\begin{aligned} U(\vec \alpha)= &  \exp[-i\vec  \alpha \cdot \vec \sigma] \\Ê
U(\vec \beta)= &  \exp[-i\vec  \beta \cdot \vec \sigma]  \end{aligned} $\\ \hline
Stokes-Mueller & $E=\begin{pmatrix} |e_x|^2-|e_y|^2 & 2 \re(e_x e_y) & 2 \im(e_x e_y) \\ 2 \re (e_x e_y^*) & -\re(e_x^2-e_y^2) & -\im(e_x^2-e_y^2) \\ -2 \im(e_x e_y^*) & \im (e_x^2+e_y^2) & -\re(e_x^2+e_y^2) \end{pmatrix}$ & $ E_\text{out}=M(\vec \alpha) E_\text{in} M(\vec \beta)^t$ & $ \begin{aligned} M(\vec \alpha)= & \exp[ 2\vec \alpha \times ] \\
M(\vec \beta)= & \exp[ 2\vec \beta \times ] \end{aligned} $\\ \hline
Four-dim.& $\vec E=\begin{pmatrix} \re(e_x)\\ \im(e_x) \\ \re(e_y) \\ \im(e_y) \end{pmatrix} $ &$\vec E_\text{out}=R \vec E_\text{in}$ 
& $ R=\exp[- \vec \beta\cdot \vec \lambda-\vec  \alpha \cdot \vec \rho]   $\\
\hline
\end{tabular}
\end{center}
\end{table*}%

\section{Discussion and Conclusions}
In Table \ref{newtable} we summarize the connection between 4d rotations, and the generalized Jones- and Stokes-Mueller-calculi, which can be seen as the main contribution of this article.
The underlying reason that 4d rotations can be modeled as 2 separate 3d rotations in this way is the isomorphism between the $SO(4)$ group and two 3d rotation groups, $O_3^+ \times  O_3^+$, as is well-known in group theory \cite{4drot}. However this is to our knowledge the first time this isomorphism have been explicitly connected to polarization analysis. 

In the introduction we mentioned three properties of these results worth emphasizing; extension, commutation, and synthesis. We will discuss them individually below.

\subsection{Extension}
There are two aspects of extension of the existing polarization calculi presented in this paper. 
The first is the mapping to real 4d geometry of the Jones calculus, which is presented here for the first time. This should be of value in communications, where a multidimensional real formalism is often preferred over complex analysis.

The second aspect is the extension of the existing Jones and Stokes-Mueller calculi to account for the left-isoclinic transformations. This were  made possible by the introduction of the state \emph{matrices} rather than state vectors. In the case of Stokes calculus this also enabled the absolute phase changes to be visualized on the PoincarŽ sphere. The idea to model absolute phase changes as rotations \emph{around} the classical Stokes vector was actually proposed previously by Frigo and Bucholtz \cite{Frigo09} (and probably understood by Pancharatnam \cite{Pancharatnam56a}) but it is interesting to see that it emerges as a natural consequence of modeling the full 4d rotation transformation set. This extension of the conventional Stokes-Mueller calculus to account for absolute phase should be quite useful, since it does not alter the conventional definitions of the Stokes vectors, but appear as a mere extension of it. In addition it preserves some of the appealing 3d properties of the Mueller calculus, by modeling birefringence as Poincar\'e sphere rotations.

\subsection{Commutation}
The fact that right- and left-isoclinic transformations commute is a unique property of the 4d rotations that greatly facilitates analysis and system realization of the forbidden rotations. Obviously it has been known, almost trivially, that absolute phase changes and polarization changes commute, but now we put it on a more firm mathematical ground also in Stokes space. The finding of the forbidden transformations, and the fact that those will commute with polarization (right-isoclinic) transformations, but \emph{not} commute with absolute phase changes is also intriguing and of potential use in future applications in the communications area.

\subsection{Synthesis}
An obvious question regarding the discovery of the forbidden transformations is: so what? If they are unphysical we need not care, right? Well, the fact that they cannot occur during photon propagation does not prevent them from being \emph{synthesized} artificially. A coherent detector followed by signal processing and a coherent transmitter will enable any transformation of the electromagnetic wave. To be able to artificially synthesize a transformation that does not affect the signal power (since it is a rotation) and that \emph{cannot be undone by wave propagation} may have interesting applications in communications, imaging, spectroscopy and signal processing. 

It is illustrative to describe a forbidden rotation in geometrical terms. The rotation described by circular birefringence is described by the 4d rotation $\exp(-\phi \rho_3)$ , which in 4d is the double rotation $D_c$ (see appendix \ref{4drot} for the exact rotation matrix) that rotates the two real and the two imaginary vector components in the same direction. The corresponding forbidden rotation $\exp(-\phi \lambda_3)$ rotates the real and imaginary parts of the two vector components in opposite directions, which is clearly unphysical for wave propagation, but not  impossible to synthesize.

\subsection{Summary and outlook}
To summarize, we have presented an extension of the conventional polarization analysis to cover the full degrees of freedom of real four-dimensional rotations. We have identified physical as well as unphysical degrees of freedom in the process. Future work on this theory could be to extend it to a full polarization calculus, accounting also for polarizing elements with polarization dependent loss or gain, as well as looking in to applications for the forbidden transformations.

\section*{Acknowledgements}
I wish to acknowledge uncountable useful and inspiring discussions with Erik Agrell on the intricacies of four-dimensional geometry. This work would not have been carried out or even initiated if it was not for those catalytic discussions. Also Colin McKinstrie should be acknowledged for many discussions on issues related (and unrelated!) to this. Colin specifically contributed with the bosonic commutation argument for the forbidden rotations used in appendix \ref{forbidden}. Helpful comments from him as well Pontus Johannisson helped improving the manuscript. Prof. Kenneth J\"arrendahl at Link\"oping University is acknowledged for sharing Ref.\cite{mueller43} with me. Financial support for this research is acknowledged from the Swedish Strategic Research Foundation (SSF) as well as from the Swedish research council (VR).

\begin{table*}[t!]
\caption{The 16 Dirac matrices $D_{jk}=\sigma_j \otimes \sigma_k$. Note that the lower right $3 \times 3$ subset  form a multiplication table, since $D_{0k}D_{j0}=D_{jk}$. The same table is given in \cite{tahir10}, and (with redefined indices) in \cite[Table 4.1]{arfken85}. \label{diractable}}
\begin{center}
{\small
\begin{equation*}
\begin{array}{|c|c|c|c|}
D_{00}= \left(
\begin{array}{cccc}
 1 & 0 & 0 & 0 \\
 0 & 1 & 0 & 0 \\
 0 & 0 & 1 & 0 \\
 0 & 0 & 0 & 1 \\
\end{array}
\right) & D_{01}=\left(
\begin{array}{cccc}
 1 & 0 & 0 & 0 \\
 0 & -1 & 0 & 0 \\
 0 & 0 & 1 & 0 \\
 0 & 0 & 0 & -1 \\
\end{array}
\right) & D_{02}=\left(
\begin{array}{cccc}
 0 & 1 & 0 & 0 \\
 1 & 0 & 0 & 0 \\
 0 & 0 & 0 & 1 \\
 0 & 0 & 1 & 0 \\
\end{array}
\right) & D_{03}=\left(
\begin{array}{cccc}
 0 & -i & 0 & 0 \\
 i & 0 & 0 & 0 \\
 0 & 0 & 0 & -i \\
 0 & 0 & i & 0 \\
\end{array}
\right) \vspace{1mm} \\
D_{10}= \left(
\begin{array}{cccc}
 1 & 0 & 0 & 0 \\
 0 & 1 & 0 & 0 \\
 0 & 0 & -1 & 0 \\
 0 & 0 & 0 & -1 \\
\end{array}
\right) &D_{11}= \left(
\begin{array}{cccc}
 1 & 0 & 0 & 0 \\
 0 & -1 & 0 & 0 \\
 0 & 0 & -1 & 0 \\
 0 & 0 & 0 & 1 \\
\end{array}
\right) & D_{12}=\left(
\begin{array}{cccc}
 0 & 1 & 0 & 0 \\
 1 & 0 & 0 & 0 \\
 0 & 0 & 0 & -1 \\
 0 & 0 & -1 & 0 \\
\end{array}
\right) & D_{13}=\left(
\begin{array}{cccc}
 0 & -i & 0 & 0 \\
 i & 0 & 0 & 0 \\
 0 & 0 & 0 & i \\
 0 & 0 & -i & 0 \\
\end{array}
\right) \vspace{1mm} \\
D_{20}= \left(
\begin{array}{cccc}
 0 & 0 & 1 & 0 \\
 0 & 0 & 0 & 1 \\
 1 & 0 & 0 & 0 \\
 0 & 1 & 0 & 0 \\
\end{array}
\right) & D_{21}=\left(
\begin{array}{cccc}
 0 & 0 & 1 & 0 \\
 0 & 0 & 0 & -1 \\
 1 & 0 & 0 & 0 \\
 0 & -1 & 0 & 0 \\
\end{array}
\right) & D_{22}=\left(
\begin{array}{cccc}
 0 & 0 & 0 & 1 \\
 0 & 0 & 1 & 0 \\
 0 & 1 & 0 & 0 \\
 1 & 0 & 0 & 0 \\
\end{array}
\right) & D_{23}=\left(
\begin{array}{cccc}
 0 & 0 & 0 & -i \\
 0 & 0 & i & 0 \\
 0 & -i & 0 & 0 \\
 i & 0 & 0 & 0 \\
\end{array}
\right)  \vspace{1mm} \\
D_{30}= \left(
\begin{array}{cccc}
 0 & 0 & -i & 0 \\
 0 & 0 & 0 & -i \\
 i & 0 & 0 & 0 \\
 0 & i & 0 & 0 \\
\end{array}
\right) & D_{31}=\left(
\begin{array}{cccc}
 0 & 0 & -i & 0 \\
 0 & 0 & 0 & i \\
 i & 0 & 0 & 0 \\
 0 & -i & 0 & 0 \\
\end{array}
\right) &D_{32}= \left(
\begin{array}{cccc}
 0 & 0 & 0 & -i \\
 0 & 0 & -i & 0 \\
 0 & i & 0 & 0 \\
 i & 0 & 0 & 0 \\
\end{array}
\right) &D_{33}= \left(
\begin{array}{cccc}
 0 & 0 & 0 & -1 \\
 0 & 0 & 1 & 0 \\
 0 & 1 & 0 & 0 \\
 -1 & 0 & 0 & 0 \\
\end{array}
\right)  \\
\end{array}
\end{equation*}
}

\end{center}
\end{table*}

\appendix
\section{The Pauli matrices} \label{paulimat}
As is common in polarization work \cite{gordon00,damask05} we define the Pauli matrices in (for physics literature, see e.g. \cite[Ch. 4.5]{arfken85}) nonstandard order, to comply with the conventional Stokes vector definition used in the classic optical text as, e.g. \cite{bow}. 
Thus the Pauli spin matrices are defined by
\begin{gather}
 \sigma_0= \left(
\begin{array}{cc}
  1 & 0    \\
  0 & 1
\end{array}
\right), \: \sigma_1= \left(
\begin{array}{cc}
  1 & 0    \\
  0 & -1
\end{array}
\right), \nonumber \\
\sigma_2= \left(
\begin{array}{cc}
  0 & 1    \\
  1 & 0
\end{array}
\right), \: \sigma_3= \left(
\begin{array}{cc}
  0 & -i    \\
  i & 0
\end{array}
\right). 
\end{gather}
which can be formed into a 3-vector with $2 \times 2$ matrices as elements $\vec {\sigma}=(\sigma_1,\sigma_2,\sigma_3)$, cf. \cite{fano54} . 
The Pauli matrices are Hermitian and have zero trace (except for $Tr(\sigma_0)=2$). They satisfy the multiplication rules
\begin{align}
\sigma_0=\sigma_1^2=\sigma_2^2= \sigma_3^2 \nonumber \\
\sigma_1 \sigma_2 =-\sigma_2 \sigma_1=i \sigma_3 \label{pauliproducts}
\end{align}
where the last equation allow for cyclic permutation of indices. This also shows that each Pauli matrix (excluding $\sigma_0$) anticommutes with the other 2 Pauli matrices, and commutes with itself and $\sigma_0$. The Pauli matrices are linearly independent and can be used as a basis for all complex $2 \times 2$ matrices. In other words, an arbitrary complex $2 \times 2$ matrix $C$ can be written as 
\begin{equation}
 C= \sum_{k=0}^3 c_k \sigma_k 
\end{equation}
where $c_k$ are complex coefficients given by
\begin{equation}
c_k=\text{Tr}(C \sigma_k)/2.
\end{equation}
If we now define $C$ as the \emph{coherency matrix }
\begin{equation}
 C= \bm e \bm e^+= \begin{pmatrix} e_x e_x^* & e_x e_y^* \\ e_x^* e_y & e_y e_y^* \end{pmatrix} \label{coherencymat},
\end{equation}
the coefficients $c_k$ are given by
\begin{equation}
c_k=\text{Tr}(\bm e \bm e^+ \sigma_k)/2=(\bm e^+ \sigma_k \bm e)/2.
\end{equation}
so we obtain the relation between Jones and Stokes vectors
\begin{equation}
 \bm e \bm e^+=\frac{P+\vec e\cdot \vec \sigma}{2}.\label{fanosformula}
\end{equation}
as originally pointed out in \cite{Falkoff51,fano54}. 

%
%

\section{The Dirac matrices} \label{diracmatrices}

The Dirac matrices $D_{jk}$ is a useful basis for $4\times 4$-matrices, and owe their name to Paul Dirac who used such matrices in his analysis of the dynamics of relativistic electrons (the Dirac equation). However, we will not follow the standard $\gamma$ notation for these matrices used in relativistic electrodynamics \cite{gammamat},  but define them as the Kronecker product of two Pauli matrices \cite[Ch 4.5]{arfken85}, i.e. as
\begin{equation}
D_{jk}=\sigma_j\otimes \sigma_k
\label{diracdef}
\end{equation}
which means that there are 16 Dirac matrices, since $j,k$ each ranges over 0,1,2,3. The 16 matrices are shown in table \ref{diractable}.

From the definition and the Kronecker product rule
\begin{equation}
(A\otimes B)(C \otimes D)=A C \otimes B D
\end{equation}
(which holds for matrices ABC and D that can be multiplied as indicated) one can derive a number of useful properties \cite[Ch. 4.5]{arfken85}:

\begin{enumerate}
\item Since the Pauli matrices are Hermitian, so are the Dirac matrices, i.e. they equal their conjugate transpose:
\begin{equation}
D_{jk}^{*t}=D_{jk}^+=D_{jk}.
\end{equation}
\item The Dirac matrices square to the unity matrix:
\begin{equation}
D_{jk}^2=\sigma_j^2\otimes \sigma_k^2=\sigma_0 \otimes \sigma_0=I.
\end{equation}
\item The Dirac matrices are traceless (except for the unity matrix), since $\text{Tr}(D_{ij})=\text{Tr}{\sigma_i}\text{Tr}{\sigma_j}=4 \delta_{i0}\delta_{j0}$.

\item The product of two Dirac matrices equals to another Dirac matrix (within a constant phase factor)
\begin{equation}
D_{jk}D_{j'k'}=(\sigma_j \sigma_{j'})\otimes( \sigma_k \sigma_{k'})
\end{equation}
where the right hand side need to be evaluated after the Pauli product rules (\ref{pauliproducts}).
\item The exponential of a Dirac matrix can be written in polar expansion form as 
\begin{equation}
\exp(i a D_{jk})=\cos(a) I+ i \sin(a) D_{jk}
\end{equation}
where $a$ is a real constant. This follows from expansion of the exponential to an infinite sum and then noting that the even terms are proportional to the unity matrix $I$ and the odd to $D_{jk}$.
\item Each Dirac matrix commutes with 8 Dirac matrices (two of these are $D_{00}$ and itself), and anticommutes with the other 8 Dirac matrices. The exception is $D_{00}$ which obviously commutes with all 16 matrices.

\item Since all Dirac matrices are linearly independent, they form a basis for all $4 \times4$-matrices. For example an arbitrary $4 \times4$-matrix $M$ can be written 
\begin{equation}
M=m_{ij} D_{ij} \label{diracexpansion}
\end{equation}
where repeated indices (here and below) indicate summation, and $m_{ij}$ are 16 complex constants. By multiplying M with $D_{kl}$ and taking the trace, only the $m_{kl}$ term remains (the others will have zero trace), and we can thus express the coefficients as
\begin{equation}
m_{kl}=\text{Tr}(M D_{kl})/4.
\end{equation}
\end{enumerate}
This last relation will help us obtain a coherency relation similar to (\ref{fanosformula}). Consider the 4d "coherency matrix" $F=\vec E \vec E^t$, where $\vec E$ is a real 4d vector. According to (\ref{diracexpansion}), $F= f_{kl} D_{kl}$ where 
\begin{equation}
f_{kl}=\text{Tr}(F D_{kl})/4=(\vec E^t D_{kl} \vec E)/4.
\end{equation}
However, since $F$ is symmetric, the 6 elements $f_{3k}$ and $f_{j3}$ with $i,j \neq3$ (corresponding to the antisymmetric Dirac matrices)
will vanish, and in addition the power is given by $4f_{00}=P$. Thus we can express the 4d coherency relation (in terms of the Stokes state matrix elements $E_{ij}$ as
\begin{equation}
\vec E^t \vec E= \frac{P+E_{ij} s_{ij}}{4} \label{4dcoherency}
\end{equation}
where $E_{ij}=\vec E^t s_{ij} \vec E$ are nine real coefficients and $s_{ij}$ denote the reordered Dirac matrices related to the Stokes state vector, defined in (\ref{smat2}) and Table \ref{diracmatoverview}. The reordering is done to have the conventional Stokes vector as the first column of the Stokes state matrix. In this way we have a useful generalization of (\ref{fanosformula}) to 4d vectors and the Stokes state matrix. 

\section{The 4d rotations} \label{4drot}
Much of the material in this appendix can be found in the Wikipedia article \cite{4drot}, but it is given here for reference and as an introduction. 

To classify the 4d rotations we start by defining the \emph{simple rotations}, that rotate only two coordinate axes while leaving the plane spanned by the other two invariant. For example a simple rotation in the 1,2-plane is 
\begin{equation}
T_{12}(\phi)=\begin{pmatrix}
 \cos(\phi) & \sin(\phi) & 0 & 0 \\
 -\sin(\phi) & \cos(\phi) & 0 & 0 \\
 0 & 0 & 1 & 0 \\
 0 & 0 & 0 & 1 \\
\end{pmatrix}
\end{equation}
which leaves the 3-4-plane invariant. Another one is
\begin{equation}
T_{24}(\phi)=\begin{pmatrix}
 1 &  & 0 & 0 \\
 0 & \cos(\phi) & 0 & \sin(\phi) \\
 0 & 0 & 1 & 0 \\
 0 & -\sin(\phi) & 0 & \cos(\phi) \\
\end{pmatrix}
\end{equation}
which leaves the 1,3-plane invariant. 
In this way we can define six simple rotations $T_{12},T_{13},T_{14},T_{23},T_{24}, T_{34}$ that span the full 6-DOF space of 4d rotations. 

The most general rotation in 4d is the \emph{double rotation}, which leaves only the origin invariant. For example we can realize a double rotation as the product of two simple rotations as
\begin{multline}
D_{a}(\phi_1,\phi_2)= 
T_{12}(\phi_1)T_{34}(\phi_2)=\nonumber \\
\begin{pmatrix}
 \cos(\phi_1) & \sin(\phi_1)  & 0 & 0 \\
 - \sin(\phi_1)& \cos(\phi_1) & 0 & 0 \\
 0 & 0 & \cos(\phi_2) & \sin(\phi_2) \\
 0 &  0 & -\sin(\phi_2) & \cos(\phi_2) \\
\end{pmatrix}
\end{multline}
And we may similarly define $D_{b}(\phi_1,\phi_2)$ and $D_{c}(\phi_1,\phi_2)$ in an analogous way as
\begin{multline}
D_{b}(\phi_1,\phi_2)= 
T_{14}(\phi_1)T_{23}(\phi_2)=\nonumber \\
\begin{pmatrix}
 \cos(\phi_1) & 0 & 0 & -\sin(\phi_1)   \\
0 & \cos(\phi_2) &  -\sin(\phi_2) & 0 \\
 0& \sin(\phi_2)  & \cos(\phi_2) &  0 \\
 \sin(\phi_1) & 0 & 0 &  \cos(\phi_1) \\
\end{pmatrix}
\end{multline}
and
\begin{multline}
D_{c}(\phi_1,\phi_2)= 
T_{13}(\phi_1)T_{24}(\phi_2)=\nonumber \\
\begin{pmatrix}
 \cos(\phi_1) & 0 & -\sin(\phi_1) & 0  \\
0 & \cos(\phi_2) & 0 &  \sin(\phi_2) \\
 \sin(\phi_1)  & 0 & \cos(\phi_1) &  0 \\
 0 & -\sin(\phi_2) & 0 &  \cos(\phi_2) \\
\end{pmatrix}
\end{multline}
to get a set of 3 double rotations that span all six DOF. 

A double rotation for which the two rotation angles are equal is called \emph{isoclinic}. We will distinguish between the $\emph{left-isoclinic}$, which have the two double rotations going in the same direction, and \emph{right-isoclinic} where the rotation angles are in opposite directions. Thus $D_{a,b,c}(\phi,\phi)$ are the left- and $D_{a,b,c}(\phi,-\phi)$ are the right-isoclinic rotations. Any simple rotation can thus be described as the product of a right- and a left-isoclinic rotation with the same angles.

The left- and right-isoclinic rotations form two 3-DOF \emph{subgroups}, i.e. any sequence of left-isoclinic rotations will always remain left-isoclinic, and vice versa. The really interesting property of the isoclinic parameterization, however, is that any left-isoclinic rotation \emph{commutes} with any right-isoclinic. Rotations within each subgroup do not commute, however. This enables any 4d rotation to be expressed as a (commuting) product of one right- and one left-isoclinic rotation. 

The underlying group theoretical reason for this is that the 4d rotation group, $SO(4)$ is isomorphic with the product of two 3d rotation groups $O_3^+ \times  O_3^+$. In other words, each of the two isoclinic groups can be mapped to the set of real 3d rotations. To present such a mapping is, in fact, the underlying purpose of this article. 

\section{Left-isoclinic photon transformations are (mostly) forbidden} \label{forbidden}
Dr. Colin McKinstrie suggested the derivation in this appendix to me.

We make the customary extension (similarly to e.g. in \cite{Bowen02,McKinstrie06}) to quantum mechanics where the electromagnetic field amplitudes $e_{x}$ and $e_y$Êcorrespond to the quantum mechanical operators $\hat \psi_{x} $ and $\hat \psi_{y}$, which obey the boson commutation relations
\begin{equation}
[\hat \psi_j, \hat \psi_k^\dagger]=\delta_{jk} \label{eq:commrel}
\end{equation}
for $j,k \in \{x,y\}$. 
We will now study input-output transformations for these operators, and the requirements the commutation relations put on the transformation matrices. We consider left- and right-isoclinic transformations separately. 
\paragraph*{Right-isoclinic transformations:}
This is the conventional input-output unitary transformation
\begin{gather}
\begin{pmatrix} \hat \psi_{xo} \\ \hat \psi_{yo}\end{pmatrix} = 
\begin{pmatrix} 
  \gamma & \eta  \\
 -\eta^* & \gamma^* 
\end{pmatrix}
\begin{pmatrix} \hat \psi_{xi} \\ \hat \psi_{yi}\end{pmatrix} \label{eq:ritransf}
\end{gather}
where subscripts i, o correspond to input, output, and the unitary property relates the complex transformation coefficients $\gamma$ and $\eta$ via
\begin{equation}
|\gamma|^2+|\eta|^2=1. \label{eq:gamdelrel}
\end{equation}
If we now apply the commutator relations (\ref{eq:commrel}) to the output operators, we find that the transformation coefficients must obey
$|\gamma|^2+|\eta|^2=1$, i.e. the commutator relations are satisfied for all right-isoclinic transformations of the polarization operators.

\paragraph*{Left-isoclinic transformations:}
This transformation is
\begin{gather}
\begin{pmatrix} \hat \psi_{xo} \\ \hat \psi_{yo}^\dagger \end{pmatrix} = 
\begin{pmatrix} 
  \gamma & \eta  \\
 -\eta^* & \gamma^* 
\end{pmatrix}
\begin{pmatrix} \hat \psi_{xi} \\ \hat \psi_{yi}^\dagger \end{pmatrix} \label{eq:litransf}
\end{gather}
where again unitarity requires (\ref{eq:gamdelrel}), but the lower-row vector components are conjugated, in contrast with the reight-isoclinic counterpart.
If we now apply the commutator relations (\ref{eq:commrel}) to the output operators of (\ref{eq:litransf}), we find that the transformation coefficients must obey
$|\gamma|^2-|\eta|^2=1$, which together with (\ref{eq:gamdelrel}) leads to 
\begin{gather}
|\gamma|=1\\
|\eta|=0,
\end{gather}
which physically corresponds to the same phase shift for both polarization components.
This means that the left-isoclinic transformation is only possible if $\eta=0$.

To conclude, we showed that the quantum mechanical commutation relations are fulfilled for right-isoclinic transformations, and left-isoclinic transformations that are pure phase shifts (i.e. have $\eta=0$ in (\ref{eq:litransf}). The remaining two DOF $(\eta \neq 0)$ for the left-isoclinic transformations are thus forbidden, or nonphysical, photon transformations.

\section{Stokes-Mueller calculus for right- and left-isoclinic rotations } \label{stokesmatrixderivation}
In the following we will sketch a derivation of the generalized Stokes calculus for the combined case of right- and left-isoclinic rotations. Our starting point is the 4d evolution equation that can be written as
\begin{equation}
\frac{d \vec E}{dz}= -K \vec E=  -( \vec h_\alpha \cdot \vec \rho + \vec h_\beta \cdot \vec \lambda)\vec E
\label{evoleq4d}
\end{equation}
where $\vec h_\alpha$ parameterizes the right-isoclinic rotation, and  $\vec h_\beta$ the left isoclinic one. The matrix $K$ is the most general antisymmetric $4 \times 4$ matrix, parameterized in terms of the vectors $\vec \rho$ and $\vec \lambda$ as defined after (\ref{rho3}) and (\ref{lambda3}). By using the definition of the Stokes matrix (\ref{smat2}) we can use (\ref{evoleq4d}) to express its differential equation as
\begin{equation}
\frac{d E_{ij}}{dz}=  -\vec E^t [s_{ij},K] \vec E
\label{evoleqsmat}
\end{equation}
for each element $E_{ij}$ in the Stokes state matrix $E$, and where $[A,B]=AB-BA$ denotes the commutator. It may now be shown (by straightforward but lengthy inspection) that the commutators of the matrices $s_{ij}$ and the rotation generators satisfy
\begin{gather}
[s_{ij},\rho_k]=-2\epsilon_{ikl}s_{lj} \\
[s_{ij},\lambda_k]=-2\epsilon_{jkl}s_{il} 
\end{gather}
where $\epsilon_{ijk}$ is the Levi-Civita tensor. We recall that the cross product between two vectors $\vec a$ and $\vec b$ can be written in tensor form 
\begin{equation}
c_i=\vec c= \vec a \times \vec b= a_j b_k \epsilon_{ijk}
\end{equation}

Thus (\ref{evoleqsmat}) can be written as an equation for the Stokes state matrix 
\begin{equation}
\frac{d E}{dz}=  (2\vec h_\alpha \times) E+ E(2\vec h_\beta \times)^t
\label{evoleqsmat2}
\end{equation}
which has the solution
\begin{gather}
 E(z)=  M(2 z\vec h_\alpha ) E(0) M(2z\vec h_\beta z\times)^t = \nonumber \\
 M(\vec \alpha) E(0) M(\vec \beta)^t
\label{evoleqsmat3}
\end{gather}
where $M$ is the 3d rotation operator defined in (\ref{orth}), $\vec \alpha = z \vec h_\alpha$ and $\vec \beta=z \vec h_\beta$. This concludes the derivation of the input-output relation for the Stokes state matrix.

\section{Relations between the Jones and Stokes state matrices} \label{statematrices}
The Jones state matrix $\bs E$ is defined with the conventional Jones vector $( \begin{array}{c} e_x \\ e_y \end{array} )$ in the first column, and its orthogonal state in the second, i.e. 
\begin{equation}
\bs E=\begin{pmatrix}
e_x &  e_y^* \\
e_y & -e_x^*
\end{pmatrix}.
\end{equation} 
The signal power is given by $P=|e_x|^2+|e_y|^2=-\det(\bs E)$.
Being a unitary matrix it can be expressed using the matrix exponential of a Hermitian matrix as follows. First define the real vector $\vec p=[\re(e_x),\re(e_y),\im(e_y)]=| \vec p | \hat p$, where $\hat p$ is the corresponding unit vector. Then we have
\begin{equation}
\bs E=(-i) \sqrt{P} \exp[ i \phi \; \hat p \cdot \vec \sigma]
\end{equation} 
where $\cos(\phi)= -\im(e_x)/\sqrt{P}$. 

We can do the same with the Stokes state matrix $E$. Being orthogonal it can be expressed as the matrix exponential of a skew symmetric matrix as
\begin{equation}
E=P \exp(-2\phi \; \hat p \times).
\end{equation}
We can also give an explicit relation between the two state matrices, namely $E_{jk} \sigma_k=\bs E^+ \sigma_j \bs E$, or more explicitly
\begin{gather}
E_{1k}\sigma_{k} =\bs E^+ \sigma_1 \bs E \nonumber \\
E_{2k}\sigma_{k} =\bs E^+ \sigma_2 \bs E \nonumber \\
E_{3k}\sigma_{k} =\bs E^+ \sigma_3 \bs E
\end{gather}
where the RHS are scalar products between $\vec \sigma$  and each of the the real row Stokes vectors $E_{1k}, E_{2k}$ or $E_{3k}$. The LHS are the product of three $2\times 2$ matrices. It is noteworthy that these relations are mathematically similar to the relation between conventional Jones matrices and 3d Mueller matrices, see e.g. \cite[Ch. 2.6]{damask05}. In terms of the vectors $[\vec e,\vec f, \vec g]$ we can write this as
\begin{gather}
(e_k,f_k,g_k)\cdot \vec \sigma=\bs E^+ \sigma_k \bs E, \hspace{1cm} k \in \{1,2,3\}
\end{gather}
where e.g. $f_k$ refers to the $k$:th vector component of $\vec f$.

 Finally, for reference it is useful to express the Stokes state matrix in terms of the Jones vector phases in normalized form, i.e. for a normalized Jones vector $\bs e=\begin{pmatrix} \cos(\theta) \exp(i \phi_x) \\ \sin(\theta) \exp(i \phi_y) \end{pmatrix}$, the Stokes state matrix becomes
\begin{widetext}
\begin{equation} 
E=\begin{pmatrix} \cos(2 \theta)  & \sin(2 \theta) \cos(\phi_x+\phi_y) & \sin(2 \theta) \sin(\phi_x+\phi_y) \\ 
\sin(2 \theta) \cos( \phi_x-\phi_y) & -\cos^2(\theta) \cos(2\phi_x)+\sin^2(\theta) \cos(2\phi_y) & -\cos^2(\theta) \sin(2\phi_x)+\sin^2(\theta) \sin(2\phi_y) \\ 
-\sin(2 \theta) \sin(\phi_x- \phi_y) & \cos^2(\theta) \sin(2\phi_x)+\sin^2(\theta) \sin(2\phi_y) & -\cos^2(\theta) \cos(2\phi_x)-\sin^2(\theta) \cos(2\phi_y) \end{pmatrix}.
\end{equation}
\end{widetext}

\bibliography{4dstokes}
\bibliographystyle{apsrev4-1}

\end{document}